\documentstyle[aps,preprint]{revtex}
\begin{document}
\draft
\preprint{\vbox{Submitted to Physical Review C
                \hfill IU/NTC 95-09}}
\title{Delta Excitations in Neutrino-Nucleus Scattering}
\author{Hungchong Kim $^1$\footnote{Email : hung@iucf.indiana.edu},
S. Schramm $^{1,2}$\footnote{Email : schramm@tpri6e.gsi.de},
C. J. Horowitz $^1$\footnote{Email : charlie@iucf.indiana.edu}}
\address{$^1$ Nuclear Theory Center, Indiana University,
Bloomington, Indiana 47408, USA \\
$^2$ GSI, D-64220 Darmstadt, Germany}
\date{\today}
\maketitle
\begin{abstract}
We derive the contribution of $\Delta$-h excitations to quasielastic
charged-current neutrino-nucleus
scattering in the framework of relativistic mean-field theory.
We discuss the effect of $\Delta$ production
on the determination of the axial mass $M_A$
in neutrino scattering experiments.
\end{abstract}
\section{INTRODUCTION}
\label{sec:intro}
The interpretation of neutrino-nucleus scattering data relies
on accurate
knowledge of the $\nu A$ cross section. Above  the quasielastic
peak higher resonances increasingly
contribute to  the cross section.
Neutrino experiments generally measure integrated yields which
include contributions from quasielastic nucleon knock-out as well as
from higher resonance production.
We discuss, in the following, the effect of $\Delta$ production
in neutrino scattering.
It has already been shown in the case of
electron scattering that a qualitative
description of the data can only be achieved when $\Delta$-h excitations are
included~\cite{wehr89,wehr93}.
Charged-current
neutrino-nucleus scattering has been used as tool to investigate axial-vector
form factor of the nucleon~\cite{ahr1}.
For low momentum transfers the $Q$-dependence of the form factor
can be parameterized by a dipole mass $M_A$, Eq.~(\ref{app:ga}).
We will consider the effect
of $\Delta$ excitations on the experimental extraction of $M_A$.
Kim {\it et al.}~\cite{hung} have examined
nuclear structure corrections to the extraction of $M_A$ but do not
consider the $\Delta$. Singh and Oset~\cite{chr} have included $\Delta$-h
but calculated in nonrelativistic formalism.

The article is organized as follows. First, in Section~\ref{sec:for}
we introduce the  relativistic
mean-field formalism of the nucleus including the $\Delta$ resonance,
deriving the appropriate set of nuclear response functions. In Section
\ref{sec:res}   we discuss the results for cross
sections and yields as modified by the inclusion of the $\Delta$ channel.

\section{Formalism}
\label{sec:for}
In this section, we derive the inclusive cross section for
quasielastic charged-current neutrino
scattering including $\Delta$-h excitations in the nucleus.
We consider a neutrino with four-momentum $k$=($E_\nu$, ${\bf k}$)
which scatters from a nucleus via $W^{\pm}$ boson exchange producing
a charged lepton  with four-momentum $k'$=($E_{\bf k'}$,${\bf k'}$).
Using an impulse approximation and a Fermi gas description of the nucleus
the formula for the double differential scattering cross section
for mass number $A$
is given by
(we assume a symmetric $N=Z$ nucleus):
\begin{eqnarray}
{d^3\sigma\over {d^2\Omega _{\bf k'} d E_{\bf k'}}}=- {A G_F^2\,
{\rm cos^2\theta _c}
\,|{\bf k}'|\over
{32 \pi^3 \rho E_\nu}}\,{{\rm Im}\,(L_{\mu\nu} \Pi_{A}^{\mu\nu})}\ ,
\label{dcross}
\end{eqnarray}
where $\rho=2k_F^3/3\pi^2$ is the baryon density with  Fermi
momentum $k_F$. $G_F$ denotes the Fermi
constant and $\theta_c$ is the Cabibbo mixing angle.
The leptonic tensor $L_{\mu\nu}$ is defined as
\begin{equation}
L_{\mu\nu} = 8 \left( k_\mu k'_\nu + k_\nu k'_\mu -
k \cdot k' g_{\mu\nu} \mp i \epsilon_{\alpha\beta\mu\nu}
k^{\alpha} k'^{\beta} \right)\ ,
\end{equation}
with the minus (plus) sign denoting neutrino (anti-neutrino) scattering.
$\Pi_{A}^{\mu\nu}$ is the polarization tensor  of the target nucleus for the
charged weak current.
Here we consider p-h, $\Pi_{ph}^{\mu\nu}$, and
$\Delta$-h, $\Pi_{\Delta h}^{\mu\nu}$,  contributions to
the polarization:
\begin{equation}
\Pi_{A}^{\mu\nu} = \Pi_{ph}^{\mu\nu} + \Pi_{\Delta h}^{\mu\nu}\ .
\end{equation}
The expressions for the p-h polarizations
have been derived in previous publications~\cite{hung,CJH}.
The weak interaction contains vector current ($v$) and
axial-vector current ($a$) contributions.
Therefore we split  $\Pi_{\Delta h}^{\mu\nu}$ into:
\begin{equation}
\Pi^{\mu\nu}_{\Delta h} = (\Pi_{\Delta h}^{vv})^{\mu\nu}
+ (\Pi_{\Delta h}^{aa})^{\mu\nu}
+ (\Pi_{\Delta h}^{va})^{\mu\nu}
+ (\Pi_{\Delta h}^{av})^{\mu\nu}\ .
\end{equation}
$(\Pi_{\Delta h}^{va})^{\mu\nu}$ and
$(\Pi_{\Delta h}^{av})^{\mu\nu}$
are interference terms of the vector and axial-vector
currents.
In a Hartree approximation the polarization tensor can be written in the form
\begin{eqnarray}
(\Pi_{\Delta h }^{ij})_{\mu\nu} =&-& i \sum_{n,p} \int{d^4p\over (2\pi)^4}\,
{\rm Tr}[\Gamma^i_{\beta\mu}(-q,-p)\ S^{\beta\alpha}(p)\
\Gamma^j_{\alpha\nu}(q,p)\  G(p-q)\ ]\nonumber \\
&+& (q_\mu \rightarrow -q_\mu)\ ~~~~~~~~~~(i,j) = (a,v)\ .
\label{eq:pol}
\end{eqnarray}
We rewrite the interference term as
\begin{equation}
(\Pi_{\Delta h}^{va})^{\mu\nu} = (\Pi_{\Delta h}^{av})^{\mu\nu} =
i \epsilon^{\mu\nu\alpha 0} q_\alpha \Pi^{va}_{\Delta h}\ .
\label{eq:mix}
\end{equation}
$S^{\mu\nu}(p)$ is the Rarita-Schwinger
form of the free spin 3/2 propagator with  momentum
$p$ \cite{wehr89}:
\begin{equation}
S^{\mu\nu}(p)=-{\not\!p + M_\Delta \over p^2 - M_\Delta^2+i\epsilon} \biggr
[ g^{\mu\nu}
-{1 \over 3} \gamma^\mu \gamma^\nu-{2 \over 3} {p^\mu p^\nu \over M_\Delta^2} +
{p^\mu \gamma^\nu - p^\nu \gamma^\mu \over 3 M_\Delta } \biggl ]\ ,
\end{equation}
Note that this expression is not unique and other forms
of the $\Delta$ propagator
have been considered (see Ref.~\cite{Muk}). The differences enter the off-shell
behavior of the propagator which do not affect
the following calculations.
The vector part of the nucleon-delta vertex has been studied
in the case of the
$\gamma N \Delta$ transition~\cite{wehr89},
\begin{eqnarray}
\Gamma^v_{\mu\nu} (q,p) &&= \sqrt{2} F_\Delta T^{\pm} \biggr
[ (-q_\mu \gamma_\nu+g_{\mu\nu}\not\!q) M_\Delta \gamma_5 +
(q_\mu p_\nu -q \cdot p g_{\mu\nu}) \gamma_5 \biggl]\ .
\end{eqnarray}
The isospin raising (lowering) operator originating from
$W^+$ ($W^-$) exchange is defined through
\begin{equation}
T^{\pm}={1\over\sqrt{2}}(T_1 \pm iT_2)
\label{eq:iso}
\end{equation}
where $2\times 4$
isospin matrices $T^i$ satisfy~\cite{tjon}
\begin{equation}
T^i (T^\dagger)^j =\delta^{ij}-{1\over 3} \tau^i \tau^j\ .
\end{equation}
The form factor $F_\Delta$ is defined in the appendix [Eq.~(\ref{app:ff})].

The vertex for the axial $N\Delta$ transition is given by~\cite{tjon,towner}
\begin{eqnarray}
\Gamma^a_{\mu\nu}&=&-r_{N \Delta}
G_A {T^{\pm} \over \sqrt{2}}
g_{\mu\nu} ~~,
\end{eqnarray}
with the axial form factor $G_A$, Eq.~(\ref{app:ga}).  The parameter
$r_{N \Delta}$
indicates the strength of axial $N\Delta$ transition and will be
discussed later.
In the noninteracting limit the nucleon propagator $G(p)$ reduces to
the free fermion propagator $G^o(p)$
for a relativistic Fermi gas with Fermi momentum $k_F$.
We consider only the
density-dependent part $G^o_F$ as vacuum contributions do not enter
at the Hartree level. In the rest frame of the nucleus
one obtains
\begin{equation}
G^o_F(p)=(\not\!p + M) {i\pi\over
E_{\bf p}}\delta(p_0-E_{\bf p}){\rm \theta}(k_F-|{\bf p}|)\ .
\end{equation}

Using an impulse approximation, the imaginary parts of $(\Pi_{\Delta h}
^{ij})$
enter the cross section Eq.~(\ref{dcross}).
As long as the $\Delta$ is assumed to be stable, the
imaginary parts
can be calculated analytically.
The resulting expressions
are given in the appendix.

In a relativistic mean field description of the nucleus,
nucleons and $\Delta$s interact with
the background of scalar ($\sigma$) and vector ($\omega$) meson mean fields.
The interactions are
assumed to be analogous to those of nucleons with possible new couplings,
$g^s_\Delta$ and $g^v_\Delta$.  The values of the couplings can be constrained
somewhat by fitting the
quasielastic peak of $\Delta$ production
in  electron scattering~\cite{wehr89,wehr93}.

In a relativistic mean-field approximation (MFA), the noninteracting
nucleon propagator $G^o_F(p)$ is replaced by
\begin{eqnarray}
G^*_F(p)=(\not\!p^* + M^*){i\pi\over
E^*_{\bf p}}\delta(p_0-E_{\bf p}){\rm \theta}(k_F- |{\bf p}|)
\end{eqnarray}
where
\begin{equation}
M^* = M-S_N~,~~
E^*_{\bf p} =\sqrt{{\bf p^2}+M^{*2}} ~,~~
p^{*\mu} =p^\mu - V_N g^{\mu 0}\ .
\end{equation}
The scalar ($S_N$) and the vector ($V_N$) self-energies can be obtained
for a given $k_F$~\cite{brian}.
Analogously, the mass and the momentum in the $\Delta$
propagator $S^{\mu\nu}$ are replaced by~\cite{wehr89}
\begin{equation}
M_\Delta \rightarrow M_\Delta^* =M_\Delta-S_{\Delta} ~,~~
t^\mu \rightarrow t^{*\mu} =t^\mu-V_{\Delta} g^{\mu 0}\ .
\end{equation}

The calculation proceeds in the same way  as in the case of the free $\Delta$.
The expressions for the
imaginary parts are given in the
appendix.     For simplicity we assume the $\Delta$ self-energies $S_{\Delta},
V_{\Delta}$ to be the same as nucleon self-energies $S_N, V_N$.

In free space, a $\Delta$ decays into $\pi$N with a width
$\Gamma=115 $ MeV.  In
the medium the situation is more complicated.  The   $\pi$N
decay channel is partially suppressed because of Pauli blocking,
i.e., the phase space
available to the nucleon produced in the $\Delta$ decay
is reduced by the Fermi sea.
However, the $\Delta$ in the medium has additional channels of decay
and obtains a
``spreading'' width with the main
decay mechanism from
$\Delta + N \rightarrow N + N$.  These two competing effects
cancel each
other partially. In our calculation, lacking better theoretical and
experimental knowledge of the $\Delta$ width in nuclear matter,
we assume a value identical to the free
width $\Gamma=115 $ MeV~\cite{wehr89,jain}.

As we are mainly interested in integrated cross sections, the results
do not depend strongly on the way the width of the $\Delta$ is
treated.  We adopt a simple method to include the decay width
by  averaging the nuclear
response over the $\Delta$ mass with a Breit-Wigner
distribution~\cite{wehr89,wehr93}.
The averaged cross section follows as
\begin{eqnarray}
\Bigl\langle {d^3\sigma\over {d^2\Omega _{\bf k'} d E_{\bf k'}}} \Bigl\rangle
&=&
\int^\infty_{M^2} d\mu^2 {d^3\sigma\over {d^2\Omega_{\bf k'}
d E_{\bf k'}}} (\mu)
f(\mu^2) ~/~ \int^\infty_{M^2} d\mu^2 f(\mu^2)~~, \\
f(\mu^2) &=&
{M_\Delta \Gamma \over (M_\Delta^2-\mu^2)^2 + M_\Delta^2 \Gamma^2 }\
\end{eqnarray}
integrating from threshold to infinity.

\section{Results}
\label{sec:res}

In this section we present result for the $\Delta$-h
calculations of charged-current neutrino interactions.
We discuss the case of muon neutrinos but the
general features of the results hold for electron neutrinos as well.
In addition to a relativistic Fermi gas calculation we consider the effects
of the mean field and include the decay width of the $\Delta$.
The target nucleus is assumed to be
$^{16}$O with a  Fermi momentum $k_F=225\, {\rm MeV} $.
For the strength of the axial $N\Delta$ transition
a simple argument using the $\Delta$ decay width suggests
a value  $r_{N\Delta} \sim 2.2$ \cite{tjon}
whereas constituent quark models give a somewhat smaller
value of $r_{N\Delta} = {6 \sqrt{2} \over \sqrt{5}} \sim 1.7$.
We choose an intermediate value $r_{N\Delta} = 2$
for our numerical calculations.

Figure~\ref{com}
shows the double differential cross section for measuring an outgoing
muon produced by an incoming neutrino
with energy $E_\nu=1$ GeV and  three-momentum transfer $|{\bf q}|=0.5 $ GeV.
First, note that the curve neglecting the decay width of the $\Delta$
has a peak around $q_0=0.37 $ GeV which agrees with the
expected elastic $\Delta$ peak at
\begin{equation}
\left(q_0\right)_{el} =\sqrt{{\bf q}^2 + M^2_\Delta} - M\ ,
\end{equation}
assuming the initial nucleon at rest.
The $\Delta$ cross section is similar to the p-h
cross section. Therefore, measurements of integrated quantities
cannot neglect $\Delta$ production. Including a finite delta width
reduces the peak height by about 30 to 35 percent but does not significantly
reduce the total integrated strength.

Nuclear matter effects are included using a
mean-field approximation (MFA). Here
we use the same scalar and vector couplings for
nucleon and $\Delta$ (known as ``universal couplings'').
For $k_F=225$ MeV, a self-consistent nuclear-matter calculation
yields the effective masses $M_\Delta^* = 931 \ {\rm MeV},$
$M^* = 638 \ {\rm MeV}$, and the vector self-energy
$V_N = V_\Delta = 239 \ {\rm MeV}$.
Mean-field results are also shown in Fig.~\ref{com}.
p-h and $\Delta$-h contributions are reduced by about
30 percent. Both peaks are shifted to higher energies due to
the smaller effective masses.

As we have seen that $\Delta$-h excitations can
contribute substantially to the charged-current cross sections,
it is interesting to study the effect of $\Delta$ production
in neutrino-nucleus scattering experiments.
Experiments using muon neutrino beams have measured
\begin{equation}
\left( d\sigma / d Q^2 \right)_{exp}
\equiv \int \frac{d\sigma}{d Q^2}(E_\nu)
f(E_\nu) dE_\nu~~.
\end{equation}
$f(E_\nu)$ denotes the spectrum of the neutrino beam.
$d\sigma / d Q^2$ is given by
\begin{equation}
{d\sigma\over dQ^2} = \int^{Q_c}_{0} {\pi \over E_\nu |{\bf k'}|}
{d^3 \sigma \over
d E_{\bf k'} d^2\Omega_{\bf k'}} dq_0
\end {equation}
where $Q^2=-q^2={\bf q}^2 - q_0^2$ and
the cut off for the energy transfer $Q_c$ reads
\begin{equation}
Q_c = E_\nu+{q^2-m^2_\mu \over 4 E_\nu }+
{E_\nu m^2_\mu\over q^2 - m^2_\mu}\ .
\end {equation}
Using the neutrino spectrum from the charged-current experiment
at BNL~\cite{ahr1} the resulting cross sections are shown in
Fig.~\ref{bnl}
where we used an axial mass $M_A = 1.09\,$GeV.
At larger momentum transfers the contribution from $\Delta$-h
excitations is as large as the nucleon knock-out.

We now discuss the influence of deltas on the extraction of the
axial mass from
quasielastic data.  Different kinds of experiments are possible.  If only a
charged lepton is detected, all $\Delta$ events will be included on an equal
footing with p-h excitations.  Alternatively, an experiment could
detect pions and thereby separate $\Delta$ events producing real pions from
p-h excitations.  However, a significant fraction of $\Delta$ excitations
lead to two-particle two-hole excitations without a real pion.  A $\Delta$ in a
nucleus can decay via $\Delta + N \rightarrow N + N$.  This is related to
either pion absorption or weak meson exchange currents (involving
an intermediate $\Delta$).  It may be difficult to separate two-particle
two-hole from one-particle
and one-hole final states.  The axial mass is often fit to reproduce the $Q^2$
dependence of observed events.  This cancels some errors from unknown flux
normalizations.   Therefore it is interesting to consider the $Q^2$ dependence
of (a) p-h excitations only (b) p-h plus that fraction (see below)
of $\Delta$-h excitations leads to 2p-2h (c) p-h plus all $\Delta$-h
excitations.
An incorrect value of $M_A$ could be extracted if one assumes
only p-h excitations while the data is ``contaminated'' by significant
$\Delta$-h excitations.

We leave it to the experimental groups to analyze their data in detail.  For
example,  Ref.~\cite{ahr1} analyzed their data assuming only p-h excitations
and extracted a value of $M_A = 1.09$ GeV with a very small statistical
error of $\pm 30$ MeV.   To estimate the uncertainty in this extracted
value of $M_A$ from $\Delta$-h excitations we try and fit the $Q^2$
dependence from 0.3 -- 1 GeV$^2$
of our full
calculation (p-h plus some fraction of $\Delta$-h) with a p-h only model.  Thus
the full calculation assumes some value of $M_A$ (which is essentially
arbitrary) and the p-h calculation attempts to reproduce this result by using a
possibly different value of $M_A$.  The important quantity is the difference
between the assumed and extracted $M_A$.  This may represent some of the
systematic error (from $\Delta$-h excitations) in a p-h only
analysis of data.

Figure~\ref{like}~(a) shows a likelihood function for reproducing our
theoretical results assuming a p-h only free Fermi gas with different
values of $M_A$.
(Note, all of the theoretical calculations used $M_A = 1.09$ GeV.)  For
theoretical calculations assuming only a p-h response (solid line),
the input $M_A = 1.09$ GeV is of course reproduced in the fit.  However,  for
theoretical calculations including either all of the $\Delta$-h (dashed line)
or half of the $\Delta$-h events (dots),  $M_A$ is underestimated by
70 to 90 MeV.

This factor of half represents a very crude estimate of the $\Delta + N
\rightarrow N + N$ to $\Delta \rightarrow N\pi$ and $\Delta + N \rightarrow
N + N$ branching ratio.  Theoretical results~\cite{had,moniz} are consistent
with this factor.
However, there could be both important $Q^2$ and model dependence in this
branching ratio.   Further theoretical work on the branching ratio
would be very useful.  Alternatively one could try and measure it
in coincidence electro-excitation experiments.

Finally in figure~\ref{like}~(b),  we fit theoretical calculations including
scalar and
vector mean fields (assumed independent of momentum) as described in
Ref.~\cite{kim}
with a free p-h calculation without mean fields.  Again large shifts
in the extracted $M_A$ are found.  $\Delta$ events tend to increase
the effective
cross section at high $Q^2$ which might be fit with a smaller $M_A$.
Alternatively,  mean field effects tend to reduce the cross section at high
$Q^2$ which can be fit using a larger $M_A$.  Thus there is some cancellation
between the two effects.  However this cancellation is unlikely to be
perfect and
the theoretical uncertainties are large.  We conclude that the theoretical
uncertainty on an extracted $M_A$ could be of order 0.1 GeV and thus large
compared to the claimed experimental error of $\pm 0.03$(stat)$\pm 0.02$(syst)
GeV~\cite{ahr1}.

\section{Summary and Outlook}
\label{sec:sum}

We have calculated charged-current neutrino cross sections including
$\Delta$-h excitations of the target nucleus.
The calculation was done for free deltas as well as including the effects
of relativistic scalar and vector mean fields in the nucleus.
$\Delta$-h excitations
are found to give significant corrections to quasielastic
nucleon knock-out processes in experiments measuring neutrinos in the
GeV range.
In extracting the axial form factor of the
nucleon from neutrino scattering data the $\Delta$-h channel
enters with similar strength as p-h contributions.
This may introduce significant error in the extracted nucleon axial form
factor.

\section*{Acknowledgments}
This research was supported in part by the DOE under Grant
No. DE-FG02-87ER-40365
and the NSF under Grant No. NSF-PHY91-08036.
\eject

\appendix
\section*{}

The imaginary parts of the $\Delta$-h
polarizations are evaluated analytically
and listed below. Note that the second term of the polarization,
Eq.~(\ref{eq:pol}), with ($q_\mu
\rightarrow -q_\mu$) vanishes for on-shell $\Delta$:
\begin{eqnarray}
{\rm Im} (\Pi_{\Delta h}^{vv} )^\mu_{\ \mu} &=& {\alpha\over 9 \pi |{\bf q}|}
F^2_\Delta
[-q^4+2 q^2 (M^2+M_\Delta^2)-(M^2-M^2_\Delta)^2 ] E^\Delta_1\ , \\
{\rm Im}(\Pi_{\Delta h}^{vv})^{00} &=&
-{\alpha\over 18 \pi |{\bf q}|} F^2_\Delta
\biggr [4 q^2 E^\Delta_3+4 q_0 \beta E^\Delta_2\nonumber \\
&&+[q^4-2 q^2
(M^2+M_\Delta^2)+(M^2-M^2_\Delta)^2+4 M^2 q_0^2]
E^\Delta_1 \biggl ]\ , \\
{\rm Im} (\Pi_{\Delta h}^{aa})^\mu_{\ \mu} &=&
{2 \alpha\over 3 \pi |{\bf q}|} G_A^2
E^\Delta_1 \ , \\
{\rm Im}(\Pi_{\Delta h}^{aa})^{00} &=& -
{2 \alpha\over 9 \pi |{\bf q}|} G_A^2
[{E^\Delta_3+2 q_0 E^\Delta_2+  q_0^2 E^\Delta_1 \over M_\Delta^2} -
E^\Delta_1]\ , \\
{\rm Im}(\Pi_{\Delta h}^{aa})^{01} &=&
-{\alpha\over 9 \pi {\bf q}^2 M_\Delta^2}
G_A^2
[2 q_0 E^\Delta_3+(3q_0^2+{\bf q}^2+M^2-M^2_\Delta) E^\Delta_2+
q_0(\beta+2 {\bf q}^2) E^\Delta_1 ]\ , \\
{\rm Im}(\Pi_{\Delta h}^{aa})^{11} &=& -{2 \alpha\over 9 \pi {\bf q}^3
M_\Delta^2} G_A^2
[ q^2_0 E^\Delta_3+ q_0 (\beta+2 {\bf q}^2)
E^\Delta_2+
({\beta^2 \over 4}+q_0^2+M^2) E^\Delta_1 ]\ , \\
{\rm Im}(\Pi_{\Delta h}^{va})&=&-{\alpha\over 9 \pi {\bf q}^2}
G_A F_\Delta [2q^2
E^\Delta_2+q_0\beta E^\Delta_1]\ ,
\end{eqnarray}
where $\alpha = [(M+M_\Delta)^2-q^2]$ and $\beta =q^2+M^2-M_\Delta^2$.
Also
\begin{eqnarray}
E^\Delta_n &=& {E_F^n-E_-^{\Delta n} \over n}~~~~(n=1,2,3)\ , \\
E_-^\Delta &=& {\rm Min}(E_F, E_{\Delta max})\ , \\
E_{\Delta max} &=& -{\beta q_0+|{\bf q}| \sqrt{\beta^2-4 M^2 q^2}
\over 2 q^2}\ .
\end{eqnarray}
${\rm Im}(\Pi_{\Delta h}^{aa})^{22}$ is obtained from the relation,
\begin{equation}
\Pi^{22}={\Pi^{00}-\Pi^{11}-\Pi^\mu_{\ \mu} \over 2}\ .
\end{equation}

In the mean field approximation, the $\Delta$ and nucleon masses in
the propagators are shifted by strong scalar fields.   Since we take the
same interaction as in
free space, the polarizations involve complicated traces.  After a
little algebra, the polarizations are written as
\begin{eqnarray}
(\Pi_{\Delta h}^{*vv})^{\mu\nu} =- {8 \over 3} F^2_\Delta \int_{M^*}^{E_F}
dE_{\bf p}
\int_{-1}^{1} d \chi \,{ |{\bf p}| \over 8 \pi^2} {T_{vv}^{\mu\nu}
\over
(p+q)^2 - M_\Delta^2 + i\epsilon  } \ ,  \\
(\Pi_{\Delta h}^{*aa})^{\mu\nu} =- {8 \over 3} G_A^2 \int_{M^*}^{E_F}
dE_{\bf p}
\int_{-1}^{1} d \chi \,{ |{\bf p}| \over 8 \pi^2} {T_{aa}^{\mu\nu}
\over
(p+q)^2 - M_\Delta^2 + i\epsilon  } \ ,  \\
(\Pi_{\Delta h}^{*va})^{\mu\nu} =- {8 \over 3} G_A F_\Delta \int_{M^*}^{E_F}
dE_{\bf p}
\int_{-1}^{1} d \chi \,{ |{\bf p}| \over 8 \pi^2} { i \epsilon^{\mu\nu\alpha 0}
q_\alpha T_{va}
\over
(p+q)^2 - M_\Delta^2 + i\epsilon  } \ ,
\end{eqnarray}
where $T_{vv}^{\mu\nu}$, $T_{aa}^{\mu\nu}$ and $T_{va}$ result
from  evaluating the traces.   These can be determined straightforwardly as
\begin{eqnarray}
T_{vv}^{\mu\nu}={1 \over 3 M_\Delta^2} \Biggr\{ &(p^* \cdot t^* -M M^*)&
\biggr[-t \cdot q M_\Delta M^*_\Delta (q^\mu t^{*\nu} + q^\nu t^{*\mu})
\nonumber \\
&& + (2 M^*_\Delta t \cdot q - M_\Delta t^* \cdot q)
(t^\mu q^\nu + p^\nu q^\mu)
M^*_\Delta
\nonumber \\
&& + (t^\mu t^{*\nu} + t^\nu t^{*\mu})
(M_\Delta^* M_\Delta q^2 -2 t \cdot q t^* \cdot q)\nonumber \\
&&-2(t\cdot q)^2 (g^{\mu\nu} M^{*2}_\Delta -t^{*\mu} t^{*\nu}) + 2 t^\mu t^\nu
[(t^* \cdot q)^2-q^2 M^{*2}_\Delta ] \biggl ]
\nonumber \\
&+ (q^\mu t^{*\nu} + q^\nu t^{*\mu})& M_\Delta \ \bigr[  M_\Delta
(M^{*2}_\Delta p^*
\cdot q + 2 t^* \cdot p^* t^* \cdot q) + t\cdot q M^{*2}_\Delta M \,\bigl]
\nonumber \\
&+M^{*2}_\Delta M_\Delta \bigr [(t^\mu q^\nu &+ t^\nu q^\mu)(t^* \cdot q M
-p^* \cdot q M^*_\Delta) + (q^\mu p^{*\nu}
\nonumber \\
&&+ q^\nu p^{*\mu}) (t^* \cdot q
M_\Delta - t \cdot q M^*_\Delta) \bigl ]
\nonumber \\
&+M^*_\Delta M_\Delta \bigr [(t^\mu t^{*\nu} + & t^\nu t^{*\mu}) (p^*
\cdot q t^* \cdot q - M M^*_\Delta q^2)
\nonumber \\
&&+ (p^{*\mu}t^{*\nu} + p^{*\nu} t^{*\mu}) ( t \cdot q t^*\cdot q -q^2
M_\Delta M^*_\Delta) \bigl]
\nonumber \\
&+2 (g^{\mu\nu} q^2-q^\mu q^\nu&) M^2_\Delta  M^{*2}_\Delta  (t^* \cdot p^*
+M^*_\Delta M)
\nonumber \\
&+ 2 t \cdot q p \cdot q M_\Delta M^*_\Delta &( g^{\mu \nu} M^{*2}_\Delta
-p^{*\mu} p^{*\nu})
\nonumber \\
&-2 t^* \cdot p^* M^2_\Delta (g^{\mu\nu} (&t^* \cdot q)^2 +q^2 t^{*\mu}
t^{*\nu})
\nonumber \\
&+2 M_\Delta M_\Delta^* t^* \cdot q g^{\mu\nu}& (t^* \cdot p^*t\cdot q
-M_\Delta
M_\Delta^* p^*\cdot q -2 t\cdot q M M^*_\Delta)
\nonumber \\
&+M_\Delta M_\Delta^* (q^2 M^{*2}_\Delta -& (t^* \cdot q)^2)(t^\mu
p^{*\nu}+t^\nu p^{*\nu}) \Biggl \}\ ,
\end{eqnarray}
\begin{eqnarray}
T^{\mu\nu}_{aa}={1 \over 6 M_\Delta^{*2}} \bigr [&& 3(t^{*2}-M_\Delta^{*2})
(p^{*\mu}t^{*\nu}+p^{*\nu}t^{*\mu})
-16 t^{*\mu} t^{*\nu} (p^*\cdot t^* + M^* M_\Delta^*)
\nonumber \\
&&+ g^{\mu\nu} t^{*2}(
p^*\cdot t^* -2 M^* M^*_\Delta)
+15 g^{\mu\nu} p^* \cdot t^* M_\Delta^{*2} + 18 g^{\mu\nu} M^* M_\Delta^{*3}
\bigl ]\ ,
\end{eqnarray}
\begin{eqnarray}
T_{va} = -{2 \over 3 M^*_\Delta} (E^*_{\bf p} - {|{\bf p}| q_0 \over |{\bf q}|}
\chi ) \bigr [&&2 p \cdot q M^*_\Delta-M^{*2}M_\Delta -4 M^* M^*_\Delta
M_\Delta
+q^2 M_\Delta\nonumber \\
&&-3 M^{*2}_\Delta M_\Delta \bigl ]\ .
\end {eqnarray}

The vector form factor of the $N\Delta$ vertex is given by
\begin{equation}
F_\Delta={-(M_\Delta+M) \over M((M_\Delta+M)^2-q^2)}\ {9\over 2}\
\biggr(1-{q^2
\over 0.71 {\rm GeV}^2}\biggl)^{-2} \biggr(1 -
{q^2 \over 3.5 {\rm GeV}^2}\biggl)^{-1/2}
\label{app:ff}
\end{equation}
which has  dipole form including some phenomenological
corrections~\cite{jones}.
For the axial vertex we use the nucleon axial form factor
\begin{eqnarray}
G_A={1.26 \over (1-q^2/M_A^2)^2}\ .
\label{app:ga}
\end{eqnarray}
\eject

\eject
\begin{figure}
\caption{Double differential cross section for
$(\nu,\mu)$ scattering from $^{16}$O showing
p-h  and $\Delta$-h contributions separately.
Results are shown neglecting (solid line) and including the
width of the delta (dashed line).
The thin lines denote  cross sections in the mean-field approximation.
}\label{com}
\end{figure}
\begin{figure}
\caption{$d\sigma/dQ^2$ averaged over BNL antineutrino spectrum are
shown for Fermi gas (solid line) and mean field approximation (dashed line).
p-h and $\Delta$-h results are shown separately.}\label{bnl}
\end{figure}
\begin{figure}
\caption{Likelihood function (normalized) of fitting experimental
antineutrino-scattering results varying the extracted axial mass, assuming a
free particle-hole (p-h) only Fermi gas response. Parts (a) and (b)
show results for Fermi-gas and mean-field approximations, respectively.
Curves are shown for p-h events (solid line), the sum
of p-h and $\Delta$-h events (dashed line), and p-h
plus half of the $\Delta$-h events (dots).  Note, all theoretical
calculations assumed $M_A=1.09$ GeV.}\label{like}
\end{figure}
\end{document}